\documentclass[12pt]{article}
\usepackage{amsmath,amssymb,theorem,cite,epsfig,url,psfrag,eepic,mathtools,amsmath}

\newcommand{\ZZ}{\mathbb{Z}}

\newcommand{\PP}{\mathbb{P}}
 \newcommand{\CC}{\mathbb{C}}

\advance \topmargin by -\headheight
\advance \topmargin by -\headsep     
\evensidemargin \oddsidemargin
\def\Maketitle{{\def\newpage{}\maketitle}}
\begin{document}
\rightline{\textit{Dedicated to the memory of Vladimir Rittenberg}}
\title{\textbf{JKLMR conjecture and Batyrev construction}\vspace*{.3cm}}
\date{}
\author{Konstantin Aleshkin$^{1,2}$, Alexander Belavin$^{1,3,4}$ and Alexey Litvinov$^{1,3,4}$\\[\medskipamount]
\parbox[t]{0.85\textwidth}{\normalsize\it\centerline{1. Landau Institute for Theoretical Physics, 142432 Chernogolovka, Russia}}\\
\parbox[t]{0.85\textwidth}{\normalsize\it\centerline{2.International School of Advanced Studies (SISSA), 34136 Trieste, Italy}}\\
\parbox[t]{0.85\textwidth}{\normalsize\it\centerline{3. Kharkevich Institute for Information Transmission Problems, 127994 Moscow, Russia}}\\
\parbox[t]{0.85\textwidth}{\normalsize\it\centerline{4. Moscow Institute of Physics and Technology, 141700 Dolgoprudnyi, Russia}}
}
\Maketitle
\begin{abstract}
We study a mirror interpretation  of the relation between the exact partition functions of $N=(2,2)$ gauged linear sigma-models (GLSM) on $S^{2}$   and K\"ahler potentials on the moduli spaces of the CY manifolds  proposed by Jockers et al. 
 We  use the Batyrev mirror construction  for establishing the explicit relation between GLSM and the  corresponding  mirror family of the Calabi-Yau manifolds, defined as hypersurfaces in weighted projective spaces.
 We demonstrate how to do this   by the explicit calculation in the case of  the quintic threefold and its mirror.
\end{abstract}
As it was shown  in \cite{Candelas:1985en} in order  to obtain the space-time supersymmetric effective theory in  $4$ dimensions  it is necessary to compactify  the  superstring theory  on Calabi-Yau (CY) manifolds.  Equivalently, it means that the compact sector of the theory has to be described by a $N=2$ superconformal field theory with the central charge $c=9$ \cite{{Gepner:1987}}. The dynamics  of the massless sector  of this theory  is governed by K\"ahler potentials of the  moduli spaces of CY manifold. Therefore  for studying the low energy dynamics one has to compute the CY moduli space geometry. 

Taking into account this aim   we study the mirror version of the  recently discovered by Jockers et al \cite{Jockers:2012dk} relation (JKLMR conjecture)  between the  K\"ahler potentials on CY moduli space and the exact partition functions of  $N=(2,2)$ gauged linear sigma-models (GLSM) on $S^{2}$ \cite{Benini:2012ui,Doroud:2012xw}. We  use the Batyrev mirror construction \cite{Batyrev:1994hm} for fixing the explicit correspondence between GLSM with  its gauge group, the set of the chiral superfields  and the corresponding mirror family of the Calabi-Yau manifolds $X$, defined as hypersurfaces in weighted projective spaces $\PP^{4}_{k_{1},k_{2},k_{3},k_{4},k_{5}}$. The key point of our approach is as follows. Let a CY hypersurface $X$ be given by zeros of the superpotential
\begin{equation}\label{fermat}
      W_X(x_{1},x_{2},x_{3},x_{4},x_{5}|\psi_{1},\dots,\psi_{h})=
         \sum_{a=1}^{h+5} C_a \prod_{i=1}^{5}x_{i}^{v_{ai}}, \quad
~  \sum_{i=1}^{5}k_{i}v_{ai}=d  \quad \text{and}~ d=\sum k_{i}.
\end{equation}
Here  $h$ is equal to the Hodge number $h_{21}$. The  coefficients $ C_a$ are some functions of  the complex structure moduli $ \psi_{1},\dots,\psi_{h}$.

In  Batyrev approach to mirror symmetry the set of the exponents  $v_{ai}$ corresponds to the lattice points of the reflexive polytope $\Delta_X$. They are  coordinates of the  vectors  $\vec V_a \in R^5$, that is $v_{ai}= (\vec V_a)_i $. 
These vectors $\vec V_a$ being  5-dimensional are subject to the linear relations
\begin{equation}\label{QV}
  \sum_{a=1}^{h+5} Q_{al} \vec V_a =0, \quad l=1,...,h,
\end{equation}
where the $ Q_{al}$ is a set of integer numbers which corresponds to the integral basis in the  linear  relations  between the exponents of the  monomials  in $W_X$. 
That is the every linear relation between $ \vec V_a$ can be expressed as a sum the relations \eqref{QV} 
with integral coefficients. This condition implies that if $\sum_{l\le h} Q_{al} m_l \in \ZZ$ for all a,
then $m_l \in \ZZ$ for all $l$.

These vectors  $\vec V_a$ are the  ``edges" of the \emph{fan}  that defines
the toric manifold $\mathbb{C}^{N}//(\mathbb{C}^{*})^{h}$, where $N=h+5$.
The symbol   $\mathbb{C}^{N}//(\mathbb{C}^{*})^{h}$  denotes the quotient  $(\mathbb{C}^{N}- Z)/(\mathbb{C}^{*})^{h}$ where  $Z$ is a certain invariant subset.
This toric manifold can be provided with a set of   projective coordinates $ y_1,...,y_N \in \mathbb{C}^{N} $, which are subject to the identification 
\begin{equation}
    (y_{1},\dots,y_{N})\sim(\lambda^{Q_{1l}}y_{1},\dots,\lambda^{Q_{Nl}}y_{N}),\quad l=1,\dots, h.
\end{equation}

Then the Calabi-Yau manifold $Y$, which is the mirror of $X$, is  realized as a subspace in the toric manifold  $\mathbb{C}^{N}//(\mathbb{C}^{*})^{h}$ defined  by the critical locus of a polynomial $W_Y(y_1,...,y_N)$ which is  weighted homogeneous with respect to all $h$ dilations 
\begin{equation}
  W_Y(\lambda^{Q_{1l}}y_{1},\dots,\lambda^{Q_{Nl}}y_{N})=W_Y(y_{1},\dots,y_{N}),\quad l=1,\dots, h.
\end{equation}
 The monomials  $$ z_j=\prod_{a=1}^{N}y_{a}^{v_{aj}},$$
 invariant with respect to the group action, can be taken 
as natural  coordinates $(z_{1},\dots,z_{N-h})$  on the toric manifold 
$\mathbb{C}^{N}//(\mathbb{C}^{*})^{h}$.

Thus $W_Y$ can be written as their linear combination
\begin{equation}\label{W-y}
W_Y(y_{1},\dots,y_{N})=\sum_{j=1}^{5} \tilde C_j \prod_{a=1}^{N}y_{a}^{v_{aj}}
\end{equation}
The manifold $Y$ given by the equation $W_Y=0$ corresponds to the reflexive polytope $\Delta_Y$ 
that is polar dual to  the polytope $\Delta_X$. Note that the matrix of the exponents in~\eqref{W-y}
is the transposed of the exponents matrix in~\eqref{fermat}~\footnote{ This connection reminds  the Berglund--Hubsch mirror construction  \cite{Hubsch1, Hubsch2}.}.

On the other hand,  it was shown by Witten \cite{Witten:1993yc} that  Calabi-Yau manifolds of such type arise as supersymmetric vacua manifolds in two-dimensional $N=(2,2)$ gauged linear sigma-models (GLSM) and  the weights $Q_{al}$ are just the charges of  $N$  chiral superfields $(\Phi_{1},\dots, \Phi_{N})$   under the gauge groups $U(1)_l$, where the subscript indicates the particular $U(1)$-component of the gauge group. Thus, knowing  the family of  CY manifolds $X$ defined as hypersurfaces in a weighted projective space by a family of polynomials $W_X$, we can find the corresponding $N=(2,2)$ gauged linear sigma-model. 
We can also find the set of its integer charges $\{Q_{al}\}$ as  such solution of  the equations \eqref{QV}
which  gives   the integral basis of the  linear  relations between the set $\vec V_a$.

Here we shortly recollect the connection \cite{Witten:1993yc}  between the supersymmetric vacua in GLSM  and the hypersurfaces in toric manifolds. Consider the  $U(1)^{h}=\prod_{l=1}^{h}U(1)_l$   $N=(2,2)$ gauge model with $h$ vector superfields and $N$ chiral matter superfields $(\Phi_{1} ,\dots \Phi_{N})$ whose   charges  under $U(1)_l$ are denoted by  $(Q_{1l},\dots,Q_{Nl})$. The  Lagrangian of the model \cite{Witten:1993yc} also depends on the coupling constants $(e_{1},\dots,e_{h})$ and on  Fayet--Iliopoulos (FI) parameters $r_{l}$, $l=1,\dots, k$,  the theta angles  and the superpotential $W_Y$.

The potential term for the scalar fields in this Lagrangian is
\begin{equation}\label{potential}
   U(\boldsymbol{\phi})=\sum_{l=1}^{h}\frac{e_{l}^{2}}{2}\left(\sum_{a=1}^{N}Q_{al}|\phi_{a}|^{2}-r_{l}\right)^{2}+\frac{1}{4}\sum_{a=1}^{N}\left|\frac{\partial W_Y}{\partial\phi_{a}}\right|^{2},
\end{equation}
where $\phi_{1},\dots,\phi_{N}$ are the scalar components of the chiral superfields  $\Phi_{1},\dots,\Phi_{N}$.  The supersymmetric ground states of the theory are parametrized by the minima of \eqref{potential} modulo gauge equivalences. For $r_{l}>0$ this space of vacua defines a manifold
\begin{equation}
  Y_{\boldsymbol{r}}=\left\{(\phi_{1},\dots,\phi_{N})\Biggl|\sum_{a=1}^{N}Q_{al}|\phi_{a}|^{2}=r_{l},\,l=1,\dots,h,\;
  \frac{\partial W_Y}{\partial\phi_{a}}=0,\right\}/U(1)^{h}.
\end{equation}
 If we send all coupling constants $e_{l}$ 
to infinity  all massive modes decouple and the model effectively reduces to the $N=(2,2)$ non-linear sigma model with the target space $Y_{\boldsymbol{r}}$.

It can be shown that the critical locus of $W_Y$ in the toric manifold $\mathbb{C}^{N}//(\mathbb{C}^{*})^{h}$ and $Y_{\boldsymbol{r}}$ are equivalent as symplectic
manifolds~\cite{Hori:2003ic}.  This construction establishes a one-to-one correspondence  between the hypersurfaces  $Y$ in toric manifolds and the gauged linear sigma-models (GLSM).

It was shown in \cite{Benini:2012ui,Doroud:2012xw} that GLSM can be  put on the 2-sphere while preserving the $N=(2,2)$ supersymmetry. The  $N=(2,2)$ supersymmetry allows one to compute the partition function of this theory exactly, using the supersymmetric  localization technique. It was done in the papers \cite{Benini:2012ui,Doroud:2012xw}. The result is given by the Mellin-Barnes type integral
\begin{equation}\label{Z-exact}
  Z_{Y}=\sum_{m_{l}\in\mathbb{Z}}\prod_{l=1}^{h}e^{-i\theta_{l}m_{l}}\int_{\mathcal{C}_{1}}\dots\int_{\mathcal{C}_{k}}\prod_{l=1}^{h}\frac{d\tau_{l}}{(2\pi i)}e^{4\pi r_{l}\tau_{l}}
   \prod_{a=1}^{N}\frac{\Gamma\Bigl(q_a/2+\sum_{l=1}^{h}Q_{al}(\tau_{l}-\frac{m_{l}}{2})\Bigr)}{\Gamma\Bigl(1-q_a/2-\sum_{l=1}^{h}Q_{al}(\tau_{l}+\frac{m_{l}}{2})\Bigr)},
\end{equation}
where the contours $\mathcal{C}_{l}$ go along the imaginary axis. The precise choice of the contours  $\mathcal{C}_{l}$ depends on the R-charges $q_a$ (see \cite{Benini:2012ui,Doroud:2012xw}).  The partition function \eqref{Z-exact} does not depend either on the coupling constants $(e_{1},\dots,e_{h})$ or on the details of the polynomial $W_Y$.  
The independence of the partition function on the coupling constants  means that we can send $e_{l}$ to infinity from the very beginning, as it was done by Witten in \cite{Witten:1993yc}. 
Thus, effectively $Z_{Y}$ picks up only the contribution of   the massless fields and gives the  exact partition function of the non-linear sigma-model.

It was conjectured by Jockers et al in \cite{Jockers:2012dk} that the exact expression for the sphere partition function \eqref{Z-exact} coincides with  the K\"ahler potential  $K_{K}^{Y}=-\log Z_{Y}$ on the quantum K\"ahler moduli space for the Calabi-Yau manifold $Y$. This conjecture was verified in \cite{Jockers:2012dk} for a few examples 
(see a physical proof in~\cite{Komar}). The problem with this check is a lack of a simple definition of the function $K_{K}^{Y}$. Therefore, instead of doing this,  we adopt a different approach based on mirror symmetry and on the Batyrev approach to it. 
 The mirror symmetry implies an equality
\begin{equation}\label{K-K-relation}
   K_{K}^{Y}=K_{C}^{X},\qquad e^{-K_{C}^{X}}=-i\int_{X}\Omega\wedge\bar{\Omega},
\end{equation}
where $K_{C}^{X}$ is the K\"ahler potential for complex structure moduli of the mirror manifold $X$. The last potential has a transparent definition in terms of integral of a uniquely (up to a function) defined holomorphic form $\Omega$ on $X$. The manifold $X$ is canonically related to the manifold $Y$ by the mirror map \cite{Batyrev:1994hm} realized as a hypersurface in toric manifold.
In these terms the conjecture  \cite{Jockers:2012dk} can be reformulated  \cite{Tanzini} as
\begin{equation}\label{Jockers-mirror}
   \int_{X}\Omega\wedge\bar{\Omega}=iZ_{Y}.
\end{equation}
We will call this the mirror version of  JKLMR conjecture.


Here we give a direct check  of the conjecture for the Quintic threefold. In this case the CY $X$ is  defined as a hypersurface in the projective space $\PP^{4}$, that is a set of five complex coordinates $(x_{1},\dots,x_{5})$ identified by
\begin{equation}
    (x_{1},x_{2},x_{3},x_{4},x_{5})\sim(\lambda x_{1},\lambda x_{2},\lambda x_{3},\lambda x_{4},\lambda x_{5}).
\end{equation}
The hypersurface $X$  $\in \PP^{4}$ is given by  zeros of the superpotential
\begin{equation}\label{ferma-2}
      W(x_{1},x_{2},x_{3},x_{4},x_{5}|\psi_{1},\dots,\psi_{h}):= 
 \sum_{a=1}^{106} C_a(\psi_{l}) \prod_{i=1}^{5}x_{i}^{\tilde v_{ai}}
   = \sum_{i=1}^{5}x_{i}^5+\sum_{l=1}^{101}\psi_{l}e_{l}(x),
\end{equation}
where
\begin{equation*}
   e_{l}(x)=\prod_{i=1}^{5}x_{i}^{s_{li}},\quad\text{with}\quad 0\leq s_{li}\leq 3, \quad h=101
\quad\text{and} \quad   \sum_{i=1}^{5} s_{li}=5.
\end{equation*}
 The parameters $\psi_{l}$ represent deformations of the complex structure of the manifold $X$.  
The manifold $Y$, which is mirror to $X$, can be realized as a subspace in the toric manifold 
defined as the quotient  $\mathbb{C}^{h+5}//(\mathbb{C}^{*})^{h}$. The action of the group 
$ (\mathbb{C}^{*})^{h}$ is given by the matrix  $Q_{al}$, as was explained above. 
Batyrev's construction tells us how to find this matrix.

Namely, the vectors of exponents $\tilde v_{ai}$ in \eqref{ferma-2}
\begin{equation}
  \tilde v_{ai}=
   \begin{cases}
        5\delta_{a,i},\quad 1\leq a\leq 5,\\
       s_{a-5,i },\quad 6\leq a\leq h+5.
   \end{cases}
\end{equation}
The vectors of exponents $\tilde v_{ai}$ are simply connected with vectors $ v_{ai}$ which 
span the one-dimensional part of the fan of the manifold $\CC^{h+5}//(\CC^*)^h$. Namely,
$$v_{ai}=\tilde v_{ai}-1.$$
As it was explained above   one  can find  the charges $Q_{al}$ of  the chiral superfields  in the corresponding GLSM   from the linear dependencies between these vectors 
\begin{equation}\label{Q}
   \sum_{a=1}^{106} Q_{al}\cdot v_{ai}  = 0.
\end{equation}
We can use in \eqref {Q} the matrix $\tilde v_{ai}$ instead of $v_{ai}$ using the additional property of the charges $Q_{al}$
$$\sum_{a=1}^{106} Q_{al}  = 0.$$

The convenient choice of solutions for this equation is 
\begin{equation}\label{Q-s}
   \tilde Q_{al}=
   \begin{cases}
       s_{la},\quad 1\leq a\leq 5,\\
       -5\delta_{a-5,l},\quad 6\leq a\leq 106.
   \end{cases}
 \end{equation}
 These relations are manifestations of the fact that any monomial can be expressed
 as a product of powers of $x_i$.

 We note, however,  that this choice of $\tilde Q_{al}$ does not give the integral basis  $Q_{al}$  in the lattice
 of all possible linear relations among $\{v_i\}_{i \le 106}$. 
That is, not all integral linear relations between $v_{ai}$ can be obtained as sums of $\tilde Q_{al}$ with
 integer coefficients. 

For instance, consider
 $s_2 = (3,2,0,0,0), \; s_3 = (2,3,0,0,0),$ which correspond to monomials
 $x_1^3 x_2^2$ and $x_1^2 x_2^3$. 
Then we see, that $\tilde Q_{a2} + \tilde Q_{a3} = (5, 5, 0, 0, 0,
 0, -5,-5,0,\ldots, 0)$ is a linear relation between the vectors $v_{ai}$. This
 relation is divisible  by $5$ and the resulting relation $(1,1,0,0,0,0,-1,-1,0, \ldots, 0)$ cannot be obtained from integral linear combinations of $\tilde Q_{al}$.~\footnote{This means that the gauge group $U(1)^{101}$ acting on the chiral fields has a discrete
 subgroup (consisting of certain fifth roots of unity) which acts trivially on all the chiral fields, which is not the case for an actual theory we are building. In other words, our choice of the charge matrix
 gives a bigger gauge group.}
 
 However, the choice of  $ \tilde Q_{al}$ instead of  $ Q_{al}$  is very convenient in computations.   
Although      $ \tilde Q_{al}$  does not give the correct charge matrix for our GLSM,
it still can be used   with one remark. 

In the formula~\eqref{Z-exact} $m_l \in \ZZ$, which is a consequence of the quantization condition for magnetic  fluxes~\cite{Benini:2012ui}. 
The actual integral basis of linear relations between $v_{ai}$ is given by the matrix $Q_{al} = \tilde Q_{ak} B^k_{l}$ for an invertible $101 \times 101$ matrix $B^k_l$.
 In the formula~\eqref{Z-exact} $m_l$ run through all possible integers. 

If we plug in the charge matrix $Q_{al} = \tilde Q_{ak} B^k_l$
in the formula~\eqref{Z-exact}, then we can define $\tilde m_l := m_k (B^{-1})^{k}_l,
\; \tilde \tau_l = \tau_k (B^{-1})^{k}_l$, the similar definition for 
$\tilde r_l := r_k B^{k}_l, \; \tilde \theta_l = \theta_k B^K_l$ and substitute the expressions
into~\eqref{Z-exact} to get rid of the matrix $B$.

 In the obtained formula the  set of all possible $\tilde m_l$ is not all integers, 
but can be described as a set of all rational numbers such that for all $a$
\begin{equation} \label{quant}
  \sum_{l \le 101} \tilde Q_{al} \tilde m_{l}\in \ZZ.
\end{equation}
This follows from  the remark after the formula~\eqref{QV}. 
Below for simplicity we will omit  the sign ``tilde'' in new notations 
 $\tilde m_l$, $\tilde V$ and $\tilde Q_{al}$.

We see from \eqref{Q-s} that $\sum_{a=1}^{106} Q_{al}=0$. Therefore, as explained in \cite {Witten:1993yc} the FI parameters and the theta parameters do not run with RG flow and the full $U(1)$ axial symmetry is unbroken. The FI and theta parameters $r_l$ and $\theta_l$ remain arbitrary parameters of the quantum (not only classical) theory.

The superpotential $W_Y$ can be written in terms of the invariant coordinates as it is presented in the formula \eqref{W-y}.  It is convenient to introduce the new notations for the $106$ chiral matter fields $(\Phi_{1} ,\dots \Phi_{106})$ and their scalar components, whose   charges  under $U(1)_l$ are given  in \eqref{Q-s},  as follows
\begin{equation}
   \Phi_{a}=
   \begin{cases}
       S_{a},\quad 1\leq a\leq 5,\\
        P_{a-5},\quad 6\leq a\leq 106.
   \end{cases}
\end{equation}
The field $P_1$ corresponds to $v_{6i}$, whose components $v_{6i}=1$  play a distinguished role. The superpotential $W_Y$ in the considered case can be written as 
\begin{equation} 
  W_Y= P_1\cdot G (S_{1},\dots, S_{5}; P_{2},\dots, P_{101}). 
\end{equation}
Being a bit sloppy, we also assign the R-charges to the fields as $q_{P_1} = 2, \; q_{P_l} = 0, \; l>1$
 and $q_{S_i} = 0$ such that the R-charge of the potential $W_Y$ is equal to 2.
The potential term for the scalar fields in the Lagrangian, which is  given by \eqref{potential} above,  will then  take the  form
\begin{multline}\label{potential-2}
   U(\boldsymbol{\phi})=\sum_{l=1}^{101}\frac{e_{l}^{2}}{2}
\left(\sum_{i=1}^{5}s_{li}|S_{a}|^{2}-5|P_l|^{2}-r_{l}\right)^{2}+   
\frac{1}{4}|G(S_{1},\dots, S_{5}; P_{2},\dots, P_{101})|^{2}+\\+
\frac{1}{4}|P_1|^{2}\sum_{i=1}^{5}\left|\frac{\partial G}{\partial S_{i}}\right|^{2}+
   \frac{1}{4}|P_1|^{2}\sum_{l=2}^{101}\left|\frac{\partial G}{\partial P_{l}}\right|^{2}.
\end{multline}

Depending on the values of FI parameters  in the theory  the different phases are expected 
to   occur \cite{Witten:1993yc , Hori:2003ic}.
In  a suitable region of $ r_l$ the vacuum manifold is a set of $ (S_i, P_l) $ obeying the equations 
\begin{equation}
  \sum_{i=1}^{5}s_{li}|S_{a}|^{2}-5|P_l|^{2}-r_{l}=0,
  \quad G(S_{1},\dots, S_{5}; P_{2},\dots, P_{101})=0, \;\; P_1 = 0
\end{equation}
  divided by the gauge group action, that is the manifold $Y$.

The exact partition function $Z_{Y}$ can be written in the form
\begin{multline}\label{main-integral}
   Z_{Y}=\sum_{m_{l} \in V}
\int_{\mathcal{C}_1}\dots\int_{\mathcal{C}_{101}}
\prod_{l=1}^{101}\frac{d\tau_{l}}{(2\pi i)}
\left(z_{l}^{-\tau_{l}+\frac{m_l}{2}}\bar{z}_{l}^{-\tau_{l}-\frac{m_{l}}{2}}\right) \times\\
\times \frac{\Gamma\bigl(1-5(\tau_{1}-\frac{m_{1}}{2})\bigr)}{\Gamma\bigl(5(\tau_{1}+\frac{m_{1}}{2})\bigr)} \prod_{a=1}^{5}\frac{\Gamma\bigl(\sum_{l}s_{la}(\tau_{l}-\frac{m_{l}}{2})\bigr)}
{\Gamma\bigl( 1-\sum_{l}s_{la}(\tau_{l}+\frac{m_{l}}{2})\bigr)}
  \prod_{l=2}^{101}\frac{\Gamma\bigl(-5(\tau_{l}-\frac{m_{l}}{2})\bigr)}{\Gamma\bigl(1+5(\tau_{l}+\frac{m_{l}}{2})\bigr)},
\end{multline}
where
\begin{equation}\label{z-theta}
   z_{l}=e^{-(2\pi r_{l}+i\theta_{l})}.
\end{equation}
The contours $\mathcal{C}$ in \eqref{main-integral} go slightly to the left of the imaginary axes: $\tau_{l}=-\epsilon+it_{l}$. The set $V$ is defined as set of all
possible $m_l$ such that $\sum_{l \le 101} Q_{al} m_l \in \ZZ$. We consider the expansion of this integral for large values of  $|z_{l}|$, that is for $r_{l}\ll0$. For $r_{l}<0$ each contour $\mathcal{C}_l$ can be closed to the right half-plane picking up the poles at 
\begin{equation}
  5\left(\tau_{l}-\frac{m_{l}}{2}\right)-1=p_{1}, \; 5\left(\tau_{l}-\frac{m_{l}}{2}\right)=p_{l};\quad p_1 = 1,2,\ldots, \; p_{l}=0,1,\dots\quad\text{such that} \quad p_{l}+ 5m_{l}>0.
\end{equation}


It is convenient to introduce the notation $\bar{p}_{l}=p_{l}+ 5m_{l}$ . Then the partition function can be rewritten as
\begin{equation}\label{res-2}
  Z=\pi^{-5}\sum_{p_1>0, p_{l}\geq0}\sum_{\bar p_{l} \in \Sigma_{p}}\prod_{l}\frac{(-1)^{p_{l}}}{p_{l}!\bar{p}_{l}!}z_{l}^{-\frac{p_{l}}{5}}\bar{z}_{l}^{-\frac{\bar{p}_{l}}{5}}
  \prod_{i=1}^{5}\Gamma\left(\frac{1}{5}\sum_{l=1}^{h}s_{li}p_{l}\right)\Gamma\left(\frac{1}{5}\sum_{l=1}^{h}s_{li}\bar{p}_{l}\right)
  \sin\left(\frac{\pi }{5}\sum_{l=1}^{h}s_{li}\bar{p}_{l}\right),
\end{equation}
where the $\Sigma_p$ is a set of all $\{\bar p_l\}$ such that $\{ \bar p_l\}$ that $\sum Q_{al} (\bar p_l - p_l)/5 = \sum Q_{al} m_l \in \ZZ$ as follows from~\eqref{quant}.
 Using the explicit expression for $Q_{al}$ the latter condition can be rewritten as $\bar p_l \in \ZZ$ and $\sum s_{li} (\bar p_l - p_l) \in 5 \ZZ$.
Also noting  that each term in \eqref{res-2}, such that $\sum_{l=1}^{101}s_{li}\bar{p}_{l}=0\;\mathrm{mod}\; 5$, vanishes, we conclude that the sum in \eqref{res-2} effectively goes over the sets
\begin{equation}
  S_{\boldsymbol{\mu}}=\left\{p_{l}, \bar p_l \; : \; \sum_{l=1}^{101}s_{li}p_{l} \equiv \sum_{l=1}^{101}
    s_{li} \bar p_l \equiv \mu_{i} \; (\mathrm{mod} \; 5)
,\;  1\leq \mu_{i}\leq 4 \right\}.
\end{equation}

Finally, using the identity
\begin{equation}
  \prod_{i=1}^{5}\sin\left(\frac{\pi }{5}\sum_{l=1}^{h}s_{li}\bar{p}_{l}\right)=(-1)^{|\boldsymbol{\mu}|}\prod_{i=1}^{5}\sin\left(\frac{\pi \mu_{i}}{5}\right)
  \prod_{l=1}^{h}(-1)^{\bar{p}_{l}},
\end{equation}

we find that
\begin{equation}\label{FM1}
  Z=\sum_{\boldsymbol{\mu}}(-1)^{|\boldsymbol{\mu}|}\prod_{i=1}^{5}
 \frac{\Gamma\left(\frac{\mu_{i}}{5}\right)}{\Gamma\left(1-\frac{\mu_{i}}{5}\right)} 
|\sigma_{\boldsymbol{\mu}}(\boldsymbol{z})|^{2},
\end{equation}
where
\begin{equation}\label{FM2}
    \sigma_{\boldsymbol{\mu}}(\boldsymbol{z})=\sum_{n_{i}\geq0} \prod_{i=1}^{5}\frac{\Gamma\left(\frac{\mu_{i}}{5}+n_{i}\right)}{\Gamma\left(\frac{\mu_{i}}{5}\right)}
   \sum_{\boldsymbol{p}\in S_{\boldsymbol{\mu},\boldsymbol{n}}}\prod_{l=1}^{101}\frac{(-1)^{p_{l}}z_{l}^{-\frac{p_{l}}{5}}}{p_{l}!}.
\end{equation}

The expression for the partition function $Z$ in \eqref{FM1} and  \eqref{FM2} coincides with the expression for 
$$ e^{-K_{C}^{X}}=-i\int_{X}\Omega\wedge\bar{\Omega},$$
 obtained by the direct computation in \cite{Aleshkin:2017fuz}. For this identification we have to  assume
that
\begin{equation}\label{z}
   z_{l}=-\psi_{l}^{-5}.
\end{equation}
This relation   is nothing but  the mirror map for the considered case.

Actually, in the region $r_l <0$, where the formula \eqref{res-2} was obtained, the theory is expected to describe the  Landau-Ginzburg phase \cite{Witten:1993yc,Hori:2003ic}. On the other hand, in order to get a metric on the K\"ahler moduli space of the manifold $Y$ which is mirror  to the Quintic one need to perform an analytic continuation to other region of $r_l$'s.  We note that the integral formula \eqref{main-integral} together with the equality \eqref{z} serve as a tool for this analytic continuation.  

In this note we  have shown how, starting from the superpotential $W(x)$ which defines a CY manifold $X$, to construct $N=(2,2)$ gauged linear sigma model whose vacuum manifold $Y$ is the mirror of $X$. 
The do this  we use Batyrev mirror construction  for fixing the explicit correspondence between GLSM and the corresponding mirror   Calabi-Yau model.
As an example of this approach we have   verified  the conjecture by Jockers et al \cite{Jockers:2012dk} for  the quintic  threefold. This paper is intended to explain
our check of the conjecture which we sketched for the class of  CY manifolds  of Fermat type in~\cite{Aleshkin:2018-2} using the results from~\cite{Aleshkin:2018jql}.

\section*{Acknowledgements}
We are grateful  to V. Batyrev, F. Benini, G. Bonelli,  D. Gepner, A. Gerhardus, H. Jockers, S. Parkhomenko, P. Putrov, A. Tanzini and F. Quevedo for the interesting discussions and  useful  comments. We are also thankful to A. Gerhardus for
pointing out a mistake in the manuscript.  A.B. is grateful to Weizmann Institute for the kind hospitality.  This work was done in Landau Institute and has been supported by the Russian Science Foundation under the grant 18-12-00439.

\end{document}